\RequirePackage{fix-cm}
\documentclass{svjour3}                     
\smartqed  
\usepackage{graphicx}
 \journalname{Hyperfine Interactions}
\begin{document}

\title{A reliable cw Lyman-$\alpha$ laser source for future cooling of antihydrogen
}


\author{D. Kolbe         	\and
				A. Beczkowiak			\and
				T. Diehl					\and
				A. Koglbauer			\and
				M. Sattler				\and
				M. Stappel				\and
				R. Steinborn			\and
        J. Walz 
}


\institute{D. Kolbe \at
              Institut f{\"{u}}r Physik, Johannes Gutenberg-Universit\"{a}t Mainz and Helmholtz Institute Mainz, D-55099 Mainz\\
              Tel.: +49-6131-39-22385\\
              Fax: +123-45-678910\\
              \email{kolbed@uni-mainz.de}           
}

\date{Received: date / Accepted: date}

\maketitle

\begin{abstract}
We demonstrate a reliable continuous-wave (cw) laser source at the 1\,$S$--2\,$P$ transition in (anti)hydrogen at 121.56\,nm (Lyman-$\alpha$) based on four-wave sum-frequency mixing in mercury. A two-photon resonance in the four-wave mixing scheme is essential for a powerful cw Lyman-$\alpha$ source and is well investigated.

\end{abstract}

\section{Introduction}
\label{intro}

Future high-resolution laser-spectroscopy of antihydrogen in a
magnetic trap can provide very stringent tests of the fundamental
symmetry between matter and antimatter (CPT symmetry)
\cite{Bluhm99}.  The two-photon 1\,$S$--2\,$S$ transition is a good candidate
for such precision experiments because of its natural linewidth of
only 1.6\,Hz at a transition frequency of 2\,466\,THz.  The absolute
frequency of this transition has already been measured in a beam with ordinary
hydrogen to enormous precision \cite{Niering00,Fischer04}.
However, the 1\,$S$--2\,$S$ ($F=1, m_F= 1 \rightarrow F=1, m_F=1$) transition
frequency has a residual dependence on the magnetic field of 186\,kHz
per Tesla.  This will broaden and shift the spectral line of
antihydrogen atoms in a magnetic trap
\cite{Cesar01}. Reducing their spatial spread in the inhomogeneous magnetic field by cooling the antihydrogen atoms will thus be very important.

Laser cooling of ordinary hydrogen atoms in a magnetic trap to the
milli-Kelvin temperature range has been demonstrated
\cite{Setija93} with a pulsed laser source at the
strong Lyman-$\alpha$ transition at 121.6\,nm wavelength from the 1S
ground state to the 2P excited state.

In addition to testing CPT there is also the intriguing prospect to
measure the gravitational acceleration of antimatter for the first
time using antihydrogen atoms \cite{Gabrielse88}. The thermal
motion of antihydrogen is a critical factor in this type of
experiments and laser-cooling at Lyman-alpha to milli-Kelvin
temperatures will be very beneficial.  Ultimately, ultracold temperatures in
the sub-milli-Kelvin range are desirable for practical experiments.
These temperatures are beyond standard laser-cooling limits for \mbox{(anti-)}hydrogen.  Novel cooling schemes for ultracold temperatures
have been proposed \cite{Walz04,Perez08,Warring09,Fischer10}.

Producing coherent radiation at 121.6\,nm (Lyman-alpha) is a challenge
as there are no tunable lasers and nonlinear frequency-doubling
crystals available for that spectral region.  Sum-frequency generation
of several incident laser beams utilizing the nonlinear susceptibility
of atomic vapors and gases is commonly used to produce coherent
radiation in the vacuum UV.  Four-wave sum-frequency mixing produces the
sum-frequency of three fundamental colors and has been
employed to generate {\em{pulsed}} laser radiation at Lyman-alpha,
typically using Krypton gas
\cite{Mahon78,Cotter79,Wallenstein80,Marangos90}.

{\em{Continuous}} coherent radiation at Lyman-alpha can have distinct
advantages for laser-cooling of antihydrogen. For example, the cooling rate is not limited by the pulse to pause ratio and the smaller linewidth reduces pumping in untrapped states. An important
difference, however, is that the power levels of continuous
fundamental beams are many orders of magnitude lower than the peak
powers typically used in pulsed Lyman-alpha generation.  Continuous
Lyman-alpha generation therefore uses resonances and near-resonances
in the nonlinear optical medium.  

In this paper a continuous coherent Lyman-alpha source is described 
which uses a four-wave mixing process in mercury with an exact two-photon resonance
and a near one photon resonance of the fundamental beams. The paper is organized as follows: First the lasersystem for the fundamental beams and the Lyman-alpha production is explained. In the next part the influence of the two-photon resonance on the four-wave mixing process is discussed. In the third section we discuss the prospect of scaling the Lyman-$\alpha$ power.

\section{Setup}
\label{sec:1}

\begin{figure*}[hbt!]
\centerline{\includegraphics[width=12cm]{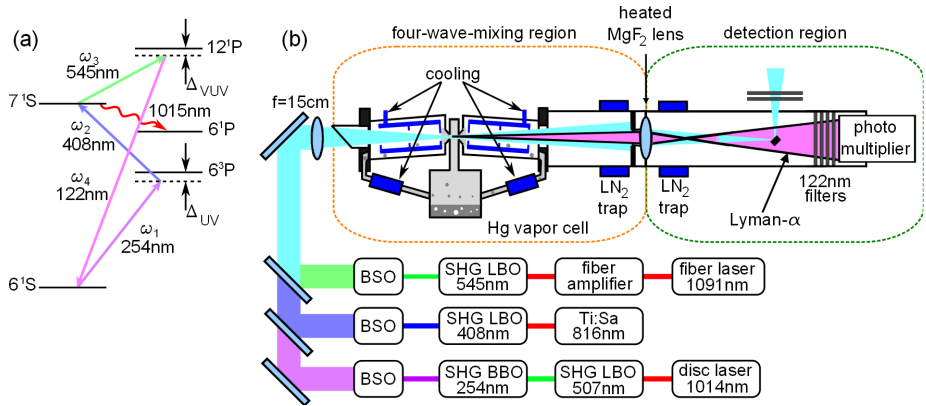}}
\caption[Level scheme] {\textbf{Energy-level diagram of mercury, the FWM scheme and setup.} \textbf{(a)}, The UV laser (254\,nm) is tuned close to the $6^1S$--$6^3P$ resonance, the blue laser (408\,nm) establishes the two photon resonance with the $7^1S$ state. With the wavelength of the green laser (545\,nm) the sum-frequency by four-wave mixing lies at the Lyman-$\alpha$ wavelength. Population in the 7$^1S$ level decays mostly (80\%) to the 6$^1P$ state at a transition wavelength of 1014\,nm.  \textbf{(b)}, Fundamental beams at 254\,nm, 408\,nm and 540\,nm wavelength are produced by frequency-doubling or frequency-quadrupling strong cw infrared solid-sate lasers. The beams are shaped, overlapped and focused into the mercury cell. In the FWM region Lyman-$\alpha$ generation takes place. Separating the Lyman-$\alpha$ from the fundamental beams is accomplished by the dispersion of a MgF$_2$ lens. Behind three Lyman-$\alpha$ filters a photomultiplier tube is used for detection. (SHG: \emph{second harmonic generation}, LBO: \emph{lithium triborate crystal}, BBO: \emph{$\beta$-barium borate crystal}, BSO: \emph{beam shaping optics}, LN$_2$: \emph{liquid nitrogen})}\label{Fig:1}
\end{figure*}

The scheme of the four-wave mixing (FWM) process is shown in Figure \ref{Fig:1}\,(a). A UV beam at 254\,nm and a blue beam at 408\,nm wavelength establish the two photon resonance between the 6$^1S$ ground state and the 7$^1S$ state of mercury. Additionally the UV beam can be tuned in a wide range around the 6$^1S$--6$^3P$ intermediate one-photon resonance. The third beam determines the Lyman-$\alpha$ wavelength and is fixed at 545\,nm.

The laser system producing the three fundamental beams is shown in the lower part of Figure \ref{Fig:1}(b). The beam at $254\,\textrm{nm}$ is produced by a frequency-quadrupled Yb:YAG disc laser (ELS, VersaDisk 1030-50). Frequency-quadrupling is done with two resonant enhancement cavities, the first one using a lithium triborate crystal (LBO) as nonlinear medium, the second one using a $\beta$-barium borate crystal (BBO). From $2\,\textrm{W}$ of infrared light at $1015\,\textrm{nm}$ we get up to $200\,\textrm{mW}$ of UV radiation. This system is in principle capable to produce up to 750\,mW of UV light, for details see \cite{Scheid07}. The second fundamental beam at $408\,\textrm{nm}$ is produced by a frequency-doubled titanium:sapphire laser (Coherent, 899-21), pumped by a frequency doubled Nd:YVO$_4$ laser (Coherent, V10). The external frequency-doubling cavity uses a LBO crystal. From $1.5\,\textrm{W}$ of IR light at $816\,\textrm{nm}$ we get up to $500\,\textrm{mW}$ of blue light. The typical day to day power of this laser system is 300\,mW. The third fundamental beam at $545\,\textrm{nm}$ is produced with a 10\,W fiber laser system at 1091\,nm (Koheras, Adjustik and Boostik) and a modified commercial frequency-doubling cavity (Spectra Physics, Wavetrain). This system is capable of producing up to $4$\,W of green light \cite{Markert07}. However, at these high powers amplified back-reflections tend to damage the entrance facet of the amplification fiber. Getting the laser repaired by the manufacturer is tedious and very time-consuming. For the present experiments we therefore operate the fiber laser at $740$\,mW, a very conservative rating, which still gives $280$\,mW of green light.   

The three fundamental beams are shaped by pairs of spherical and cylindrical lenses (see Figure \ref{Fig:1}(b)). The beams are overlapped at dichroic mirrors and focused into the mercury cell using a fused silica lens with a focal length of 15\,cm. The mercury cell can be heated up to 240\,$^\circ$C providing a mercury vapor density of up to $N=1.1\times 10^{24}\textrm{m}^{-3}$. Outside the focus region cooled baffles and helium buffer gas with a pressure of 100\,mbar are used to avoid condensation of mercury on the optics. The four-wave mixing region is separated from the detection region by a vacuum sealed MgF$_2$ lens which performs separation of the Lyman-$\alpha$ light from the fundamental beams (see Figure \ref{Fig:1}(b)). Due to the dispersion of this lens the focal length differs for the Lyman-$\alpha$ wavelength ($f=21.5$\,cm at 540\,nm, $f=13$\,cm at 122\,nm). A tiny mirror is placed in the focus of the fundamental beams to reflect them to the side. The Lyman-$\alpha$ beam is large at the fundamental focus and therefore the mirror just casts a shadow in the Lyman-$\alpha$ beam, causing $\approx 30\%$ loss. A solar-blind photomultiplier tube is used for detection of the Lyman-$\alpha$ photons. Background is suppressed by three 122\,nm filters. The overall detection efficiency due to losses in the MgF$_2$ lens, the tiny mirror, the three filters and the photomultiplier efficiency is $6 \times 10^{-5}$. IR light from the 7$^1S$--6$^1P$ decay at 1014\,nm (see also Figure \ref{Fig:1}(a)) can be detected as a measure of the beam overlap of the blue and the UV beam.

\section{Two-photon resonance}
To enhance the third order nonlinear susceptibility exploition of a two-photon resonance has been investigated \cite{Smith88}. For a powerful cw Lyman-$\alpha$ source the utilisation of a two-photon resonance is essential \cite{Eikema99,Walz01,Scheid09}. We use the $6^1S$--$7^1S$ two-photon resonance with a transition wavelength of 156.5\,nm. The nonlinear susceptibility $\chi^{(3)}$ responsible for the FWM in the case of a close two-photon resonance can be expressed as \cite{Smith87}:

\begin{equation}
\chi^{(3)}=\frac{N}{6 \epsilon_0 \hbar^3} S(\omega_1,\omega_2) \chi_{12} \chi_{34}
\label{chi3}
\end{equation}

with the two partial susceptibilities 

\begin{equation}
\chi_{12}= \sum_m\left(\frac{ p_{nm} p_{mg} }{\omega_{gm}-\omega_1}+ \frac{p_{nm} p_{mg} }{\omega_{gm}-\omega_2}\right)
\label{chi12}
\qquad ,
\end{equation}

\begin{equation}
\chi_{34}= \sum_\nu \left(\frac{ p_{n\nu} p_{\nu g} }{\omega_{g\nu}-\omega_4}+ \frac{ p_{n\nu} p_{\nu g} }{\omega_{g\nu}+\omega_3}\right)
\label{chi34}
\qquad ,
\end{equation}

and the term describing the two-photon resonance

\begin{equation}
S(\omega_1+\omega_2)=\frac{1}{\omega_{ng}-(\omega_1+\omega_2)}
\label{Glng:Seinfach}
\qquad .
\end{equation}

Identical linear polarization of the fundamental beams is assumed. Summing over $m$ and $\nu$ in the partial susceptibilities includes all exited states that connect to the $6^1S$ ground state (index n) and the $7^1S$ state (index g) by dipole transitions. The dipole matrix elements $p_{ab}$ can be obtained from the oscillator strengths $f_{ab}$ tabulated in \cite{Alford87}. The function $S(\omega_1+\omega_2)$ contains the enhancement due to the two-photon resonance. The homogeneous line-broadening $\Gamma^{hom}_{7S}$ consisting of natural line width and pressure-broadening is included by adding the term $-i\Gamma^{hom}_{7S}/2$ in the denominator of equation (\ref{Glng:Seinfach}). The Doppler-broadening is included by adding the Doppler-shift $kv$ of an atom with velocity $v$ and integrating over the one dimensional Boltzmann velocity distribution. The two-photon resonance has two effects: On the one hand, the two-photon resonance enhances the FWM process according Equation \ref{chi3}. On the other hand, it transfers population to the $7^1S$ two-photon state which can be detected with the fluorescence light of the 7$^1S$--6$^1P$ transition at 1015\,nm (see Figure \ref{Fig:1}(a)). In the next two subsections we will first discuss the effect of polarization of the fundamental beams on the FWM process and the population transfer and second introduce a laser process on the 7$^1S$--6$^1P$ transition as a adjustment tool.

\subsection{Polarization dependence}
\label{sec:2}

\begin{figure*}[hbt!]
\centerline{\includegraphics[width=7cm,angle=-90]{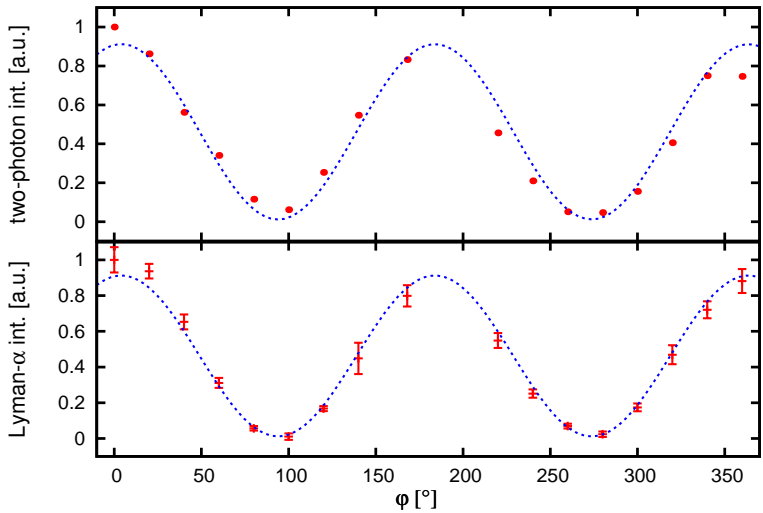}}
\caption[Level scheme] {\textbf{Polarisation dependence of the IR and Lyman-$\alpha$ signal.} Normalized IR (top) and Lyman-$\alpha$ (bottom) intensity versus the angle $\varphi$ between the polarization of the blue and UV beam. Both signals are modulated in the same way following a cosine function. Periode and phase of the function is fitted to the Lyman-$\alpha$ signal.}\label{Fig:2}
\end{figure*}

In Equation \ref{chi3} we assumed equal polarization of all fundamental beams. In general the third order nonlinear susceptibility is a tensor of the order of 4 including all possible polarization states of the fundamental beams. In the case of two-photon resonant four-wave mixing the efficiency depends critically on the polarisation of the UV and blue beams which establish the two-photon resonance \cite{Irrgang98}. In Figure \ref{Fig:2} the normalized IR signal of the decay of the $7^1S$ and the normalized Lyman-$\alpha$ signal out of the four-wave mixing process versus the angle between the linear polarization of the UV beam to the blue beam is shown. It can be clearly seen that both signals are equally modulated by the polarisation. In the case of the two-photon signal excitation to the $7^1S$ state can only occur with equal polarization of the fundamental light fields. The excitation is therefore modulated by the intensity of the UV beam projected on the polarisation vector of the blue beam resulting in a modulation proportional to a cosine function. The same behavior is observed in the Lyman-$\alpha$ signal showing that the two-photon resonance in the nonlinear susceptibility depends of the polarisation in the same manner. The phase of the cosine function in Figure \ref{Fig:2} is extracted from the Lyman-$\alpha$ signal. The polarisation of the green beam in contrast did not have any influence on the four-wave mixing efficiency as was checked by rotating the green polarization and has already been well investigated \cite{Irrgang98}. 

\subsection{Two-photon laser induced stimulated emission}

\begin{figure*}[hbt!]
\centerline{\includegraphics[width=7cm,angle=-90]{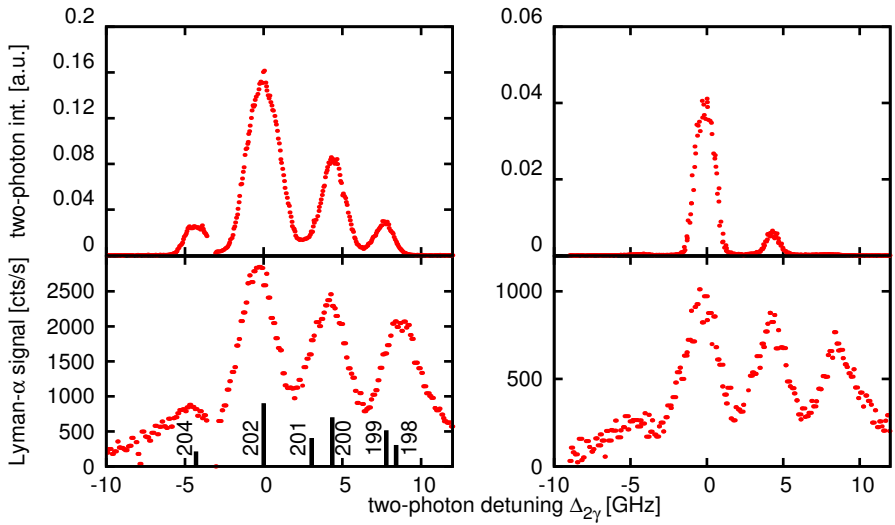}}
\caption[Level scheme] {\textbf{Alignment dependence of the IR and Lyman-$\alpha$ signal.} IR (top) and Lyman-$\alpha$ (bottom) power at scans across the two-photon resonance. Two-photon detuning $\Delta_{2\gamma}=\omega_{6^1S-7^1S}-(\omega_1+\omega_2)$ determines the detuning to the two-photon resonance of the 202 mercury isotope. At threshold of the TALISE process the IR power is much more critical on the alignment of the beams then the Lyman-$\alpha$ power. Power increase from bad alignment (right) to good alignment (left) differs for the IR and Lyman-$\alpha$ signal. Vertical bars indicate the positions and abundances of the different isotopes. Note the different scales. 
}\label{Fig:3}
\end{figure*}

Due to the short lifetime of the $6^1P$ state (1.5\,ns) compared to the two-photon $7^1S$ state (32\,ns) a high enough two-photon excitation rate can establish a laser process on the $7^1S$--$6^1S$ transition at 1015\,nm (see Figure \ref{Fig:1}(a)). This two-photon laser induced stimulated emission (TALISE) has been well investigated \cite{Amorim01,Kolbe10b}. It does not influence the four-wave mixing process in this low loss (due to conversion) regime. Due to the strong increase in the IR power at the threshold condition of the TALISE process the IR signal is a perfect diagnostic tool for the overlap of the blue and UV beam. In Figure \ref{Fig:3} the IR signal and Lyman-$\alpha$ versus the detuning to the two photon resonance is shown. At the bottom the Lyman-$\alpha$ signal peaks at the two-photon resonance condition of every mercury isotope resulting in four peaks. The threshold condition for TALISE depends of the density of the mercury atoms and therefor differs for the various isotopes. At the right of Figure \ref{Fig:3} the threshold condition of the 202 and 200 isotopes (the most abundant isotopes in a natural mercury mixture \cite{Zadnik89}) are fullfilled resulting in two peaks in the IR signal (top). The IR signal is at threshold much more critical on the alignment of the overlap of the blue and the UV beam. Better alignment (left) boosts the IR signal to a greater extent than the Lyman-$\alpha$ signal which provides a much more sensitive adjustment.

\section{Lyman-$\alpha$ power scalability}

In the case of four-wave mixing with no loss of the fundamental beams due to absorption and conversion the power at Lyman-$\alpha$ is proportional to the power-product of the fundamental beams \cite{Bjorklund75}:

\begin{equation}
 P_{4} \propto  P_1 P_2 P_3 
\qquad ,
\label{Glng:BjorklundG}
\end{equation}

where $P_4$ is the power at Lyman-$\alpha$ and $P_{1,2,3}$ is the power of the UV, blue and green fundamental beam respectively. The Lyman-$\alpha$ power was measured to be perfectly linear to the green power. We achieved a maximum power of 0.3\,nW at fundamental powers of $P_{1}=130$\,mW, $P_{2}=324$\,mW and $P_{3}=185$\,mW. The power levels required for Doppler-cooling of antihydrogen in a magnetic trap are rather low: With $1\,$nW of Lyman-$\alpha$ we expect cooling times in the order of minutes \cite{Walz01}. Nevertheless, due to losses at the beam transfer from the four-wave mixing region to the antihydrogen trap center and for reducing of cooling time a high power Lyman-$\alpha$ source is desired. With full fundamental laser powers we expect Lyman-$\alpha$ powers up to 65\,nW. Further increase could be performed by enhancing the fundamental powers in separated cavities with a shared focused cavity arm where all beams are overlapped and FWM takes place. With typical coupling efficiencies of $\approx$70\% and estimated enhances of a factor of 50 for every fundamental beam an increase in Lyman-$\alpha$ power of a factor of 40\,000 seems feasible.

\section{Conclusion}

A cw Lyman-$\alpha$ laser source is a key tool for future experiments on antihydrogen. We presented a reliable Lyman-$\alpha$ source with powers up to 0.3\,nW for an estimated cooling time of antihydrogen in the region of minutes. The $6^1S$--$7^1S$ two-photon resonance in mercury is crucial for the four-wave mixing efficiency.

%

\begin{acknowledgements}
We acknowledge support from the Bundesministerium f\"{u}r Bildung und Forschung.
\end{acknowledgements}

\bibliographystyle{spphys}       
\bibliography{bib}   

%
%

\end{document}